\documentclass[amsfonts,amssymb,aps,draft,floatfix,twocolumn]{revtex4}

\usepackage{epsfig}

\def\beq{\begin{equation}}
\def\eeq{\end{equation}}
\def\2{\mbox{$1\over2$}}
\def\6{\langle}
\def\9{\rangle}

\newcounter{step} 
\newcommand{\debprotocol}[1]{
\begin{center}{\bf #1}
\end{center}
\begin{list}{}{
\setlength{\leftmargin}{0pt}
}}

\newcommand{\finphase}{\end{list} \mbox{}}
\newcommand{\finprotocol}{\end{list} \setcounter{step}{0}}

\newtheorem{definition}{Definition}
\newtheorem{proposition}{Proposition}

\newtheorem{theorem}{Theorem}

\newcommand{\blue}[1]{#1}
\newcommand{\green}[1]{#1}

\newcommand{\debproof}[1]{\par \addvspace{7pt} \noindent {\em Proof{#1}.} }
\newcommand{\finproof}{\mbox{$\hfill \Box$} \par \addvspace{7pt} \noindent}

\newcommand{\spec}{\mbox{$(P^A_a, P^B_b, \Phi^+_{AB})$}}

\newcommand{\realD}{\mbox{$(P^{\tilde B}_b, P^{\tilde A}_a,
\Psi_{\tilde A \tilde B})$}}

\newcommand{\realP}{\mbox{$(P^{\tilde B}_b, P^{\tilde A}_a,
\Psi_{\tilde A \tilde B})$}}

\begin{document}
\title{Self testing quantum apparatus} 
\author{Dominic Mayers$^*$ and Andy Yao$^{**}$} 
\affiliation{* Caltech, Pasasena, USA\\
** Princeton University, Princeton, USA}

\date{\today}

\begin{abstract}
We study, in the context of quantum information and quantum
communication, a configuration of devices that includes (1) a source
of some unknown bipartite quantum state that is claimed to be the Bell
state $\Phi^+$ and (2) two spatially separated but otherwise unknown
measurement apparatus, one on each side, that are each claimed to
execute an orthogonal measurement at an angle $\theta \in \{-\pi/8, 0,
\pi/8\}$ that is chosen by the user. We show that, if the nine
distinct probability distributions that are generated by the self
checking configuration, one for each pair of angles, are consistent
with the specifications, the source and the two measurement apparatus
are guaranteed to be identical to the claimed specifications up to a
local change of basis on each side.  We discuss the connection with
quantum cryptography.
\end{abstract}

\maketitle

\section{Introduction}
Typically, when one considers the task of testing a quantum system,
for example using quantum state tomography, one makes the assumption
that the measurement apparatus are perfect or reasonably close to
perfect. Moreover, it is typically assumed that every measured system
has the correct dimension.  In contrast, here we consider the problem
of testing a quantum system without trusting the measuring apparatus
that are used in the test, except the fact that two measurements that
are space like separated in the ideal specification can be modeled in
the real setting by two silmutaneous quantum operators on distinct
systems. In particular, we assume no apriori information about the
dimension of the measured systems or on the rank of the measurement
operators.  The intuition that applies to ordinary tests where the
dimensions are correct does not apply here.  The problem that we have
is more difficult. In fact, it will already be an interesting
challenge to consider the case where the probability distribution for
measurement outcomes in the real setting are identical to the
corresponding probability distributions in the ideal specification.

To be more specific, we describe the setting that we will
consider. The source has specification to emit two systems (say
photons) $A$ and $B$ in the Bell state $\Phi^+ = (|00\rangle^{AB} +
|11\rangle^{AB} )/\sqrt{2}$.  The photons $A$ and $B$ are sent to two
measuring apparatus, one for each photon, that respectively receive
the classical inputs $\alpha, \beta \in \{-\pi/8, 0, \pi/8\}$ that
represent the measurement bases $\{\, |\alpha + 0\rangle,\, |\alpha +
\pi/2 \rangle\, \}$ and $\{\, |\beta + 0\rangle,\, |\beta + \pi/2
\rangle\, \}$.  The photons $A$ and $B$ are respectively measured in
the bases $\alpha$ and $\beta$ and the respective classical outcomes
$x$ and $y$ of these measurements are noted on each side.

Let $p((\alpha,x),(\beta,y))$ be the probability of the pair of
outcomes $(x,y)$ given the pair of measurements $(\alpha,\beta)$ when
the system respects perfectly the original specification.  Let $\tilde
p((\alpha,x), (\beta,y))$ be the corresponding probabilities for the
actual system which might not be built to the orginal specification.
Let $\Theta = \{ -\pi/8, 0, \pi/8\} \times \{0,1\}$, the space of
pairs $(\alpha, x)$ where $\alpha$ is a basis and $x$ an associated
outcome. Our main result states that if, $\forall (a, b) \in
\Theta^2$, we have $\tilde p(a,b) = p(a,b)$, the setting is
neccessarily identical modulo some local isomorphisms (see section
\ref{Problem}) to the original specification.  This result would not
be surprising at all if we assumed that the two measured systems are
two dimensional systems or that the measurement operators are executed
on two dimensional systems, but we do not use any assumption of this
kind here.  No assumption on the measuring apparatus or the source are
required in our proof of the theorem, except that before a measuring
apparatus receives its choice of basis and until after the measurement
is executed, the apparatus (and whatever has selected the basis) is
isolated from the other measuring apparatus. In the specific context
of quantum key distribution, this separation assumption is also needed
after the measurement to guarantee the privacy of the generated key,
but this is another issue. If we only worry about testing the source,
we only need the separation assumption until after the test is
executed.  This separation assumption is only an assumption on the
device or mechanism that is used to isolate the measuring apparatus;
it is not a direct assumption on the measuring apparatus.

If the actual system is not built to the original specification, we
might have $\tilde p(a,b) \neq p(a,b)$. A test could eventually be
executed to check how close the probabilities $\tilde p(a,b)$ are to
the ideal case $p(a,b)$.  A robust variation on the theorem should
consider the case $|\tilde p(a,b) - p(a,b)| \leq \epsilon$, but we do
not do this analysis.  The problem and the main theorem are described
in section \ref{Problem}. A specific connection with the BB84 protocol
\cite{bb84} in quantum cryptography is discussed in section
\ref{BB84Connection}. Finally, the proof of the theorem is provided in
section \ref{Proof}.
 
\section{Related results} \label{Context}
The result is interesting from both a purely theoretical point of view
and a practical point of view. We hope that it will have application
in different areas of quantum information processing.  The result was
obtained in a specific context, the unconditional security
\cite{mayers96} of a variation on a protocol proposed by Bennett and
Brassard in 1984 \cite{bb84}.  For concreteness, we first explain this
(variation on the) protocol.  In this protocol, Alice sends many
photons to Bob in one of the four polarisation states $|\alpha + x
\pi/2\rangle$, $\alpha \in \{-\pi/8, \pi/8\}$, $x \in \{ 0,1\}$.
Usually, in the literature, the bases are at angles are $0$ and
$\pi/4$ instead of $-\pi/8$ and $\pi/8$ as we have here, but this is
symmetrical.  Bob measures each received photon in one of these two
bases chosen uniformly at random and notes the outcome $y$. Let
$\Omega$ be the set of positions where Alice and Bob used the same
basis.  For each postion in $\Omega$, because Alice and Bob used the
same basis, in principle they should share the same bit $x = y$.
After the quantum transmission, Alice and Bob announce their bases,
and therefore they learn the set $\Omega$. Eve also learns the bases
used by Alice, but it is too late for Eve because the photons are
already on Bob's side.  Alice and Bob execute a test on a subset $T
\subseteq \Omega$ and count the number of errors in $T$. If too many
errors are detected the test fails and the protocol aborts. Otherwise,
the set $T$ is thrown away and the protocol continues.  Alice
announces redundant bits about $E = \Omega - T$ and Bob uses this
extra information to correct the errors in $E$.  At this point, Alice
and Bob should share a string of bits on $E$ which we call the raw
key.  Eve has obtained some information about the raw key from the
redundant information or directly because she eavesdropped on the
quantum channel.  To address this problem, Alice and Bob generate a
final key ${\bf k}$ in which each bit $k_i$ is the parity of some
subset $K_i \subseteq E$ of the raw key bits, a well known technique
to extract a final key from the raw key \cite{bbr88}.  As proven in
\cite{mayers96}, if the number of parity bits is not too large, the
final key will be almost perfectly secret, that is, with almost
certainty either the test fails or Eve has almost no information about
the final key.  The connection between this proof and our result is
given in section \ref{BB84Connection}.

The novelty of the problem that we consider is that the apparatus that
are used in the protocol can be defective.  There is no perfect
solution to this problem.  Therefore, every tool that can be used to
improve the situation is normally welcome and our theorem is such a
tool.  Before we explain in which way our theorem helps, let us
describe the basic steps that should be taken even before we use this
theorem.  The first basic step is simply to propose assumptions that
directly say that the source is close to the ideal specification. The
second basic step is to justified each of these assumptions (as much
as an assumption can be) by considering the specific technology used.
An example of such a direct assumption is an upper-bound on the ratio
of multi-photon signal that is emitted by a source.  In this case, the
required technology could be a strong signal of light followed by an
attenuator. The intensity of the strong signal is measured and then
the attenuator is used to reduce the intensity of the signal to the
desired level. It turns out that this specific technology is
reasonably trusted.  In the same way, a direct assumption must be
proposed and justified for every degree of freedom which can encode
information.  In addition to the photon number space, one has to
consider the frequency space, the polarization space and also we must
consider the possibility that the information gets encoded into a
different system. For example, if any mechanical system is used, the
information could get encoded into some vibrational modes.  Moreover,
one should explain (in the general model of quantum mechanics) why all
aspects are covered by the proposed assumptions. So the task is not
easy and it is not a perfect solution.

Some people might say that these assumptions should not be trusted,
and a test should be conducted to verify them instead.  For example,
one might propose to use a photon detector to test the intensity of
the signal after the attenuator. Fine, but this leads us to our
complementary approach. The immediate problem that we must address is
that validity of the test depends on the testing apparatus, the photon
detector.  Let us consider what can be done. In 1991, Artur Ekert
proposed a protocol that was based on EPR pairs and violation of a
Bell inequality \cite{ekert91}. Though the security analysis of his
protocol was incomplete, Ekert`s salient idea of using EPR pairs and
violation locality in a protocol is far reaching.  First, if the
parties on both sides have trusted measuring apparatus, the idea of
using an untrusted source of EPR pairs takes care of the defective EPR
source issue by itself (without any use of violation of locality).
Ekert's analysis considered the case in which we have perfect
measuring apparatus.  A little bit later, it was strongly suggested in
\cite{bbm92} that violation of a Bell inequality (which is more than
just using EPR pairs) does not help in quantum cryptography unless the
purpose is to verify the quantum apparatus.  In the context of trusted
quantum apparatus, violation of Bell like inequalities might not
provide better security. Moreover, to our knowledge, it didn't help as
a theoretical tool to obtain a better bound on Eve's
information. However, with violation of locality, one can hope to test
both a defective source and defective measuring apparatus.  Ekert did
not explicitly propose that we should use violation of locality to
test both the EPR source and the measuring apparatus, but the idea was
implicitly there.  It was there, but remained unused until the work of
\cite{my98} in which violation of locality was used to test both the
source and the measuring apparatus (see also \cite{mt00}).  Two papers
reported experimental work that was based upon Ekert's protocol (or a
similar protocol)~\cite{npwb00,jswwz99}.  The purpose of some these
papers was to address the imperfect apparatus issue.  They cite Ekert
for his protocol.  However, nobody knows if Ekert's protocol meets the
objective.  Even Ekert did not claim that.  His main focus was not to
test defective apparatus.  There are other connections between quantum
cryptography and violation of locality or Bell inequalities that focus
less on testing defective apparatus than the current paper does, but
are not less interesting \cite{gisin02}.

This paper, following \cite{my98}, shows that, except for the location
of each apparatus and the time at which they receive their input, we
can assume that Eve has designed both the EPR source and the
detectors.  In practice, to conduct a test on the source, we also need
the assumption that the different executions of the overall
configuration in the test are identical and independent.  Our main
tool is violation of classical locality. In the context of untrusted
source and untrusted measuring apparatus, it is not clear whether or
not a violation of a Bell inequality, especially if it is only
sligthly violated, implies that a private key can be generated.
Therefore, in this context, it is not sufficient to detect a violation
of a Bell inequality to claim that a protocol is secure. So, our
statement is not that violation of locality implies security. We use
violation of locality indirectly by considering an ideal setting that
is known to violate a Bell inequality and by restricting ourselves to
attacks that generate the same probability distribution of classical
outcomes as in this ideal setting.  Again, if the attack modifies the
probability distribution of these classical outcomes, this can be
detected by Alice and Bob, but we do not do this part of the analysis.
Of course, this test will require its own set of assumptions.  This
approach should not be taken as a way to avoid the basic steps that
are described above. Instead, in view of the fact that no solution is
perfect, we think that the existence of two complementary approaches
is very much welcome.

\section{The problem and the main result}\label{Problem}
An apparatus emits an unkown system $\tilde A \otimes \tilde B$ in
some unknown state $\Psi_{\tilde A \tilde B}$ which will be measured
by some unknown measurement operators $\Pi^{\tilde A}_a$, $\Pi^{\tilde
B}_b$ associated with $a, b \in \Theta$, respectively.  Without loss
of generality, we assume that the source emits a pure state
$\Psi^{\tilde A \tilde B}$. There is no loss of generality because an
extra system can be added to $\tilde A$ or $\tilde B$ to purify the
state.  Without loss of generality, we can further assume that the
defective measurement operators $\Pi^{\tilde A}_a$ and $\Pi^{\tilde
B}_b$ respectively associated with $a$ and $b$ are orthogonal. An
auxiliary system in some known state can be added to the measured
system to replace a general POVM on this system by an orthogonal
measurement on the extended system \cite{neumark43}.  The ideal
projection on the system $A$ ($B$) associated with an element $a \in
\Theta$ ($b \in \Theta$) is denoted $P^A_a$ ($P^B_b$).  We define
$p(a,b) = \| P^A_a \otimes P^B_b |\Phi^+_{AB} \rangle \|^2$ and
$\tilde p(a,b) = \| \Pi^{\tilde A}_a \Pi^{\tilde B}_b | \Psi_{\tilde A
\tilde B} \rangle \|^2$.

As we explained in the Introduction, the hypothesis in our main result
(theorem \ref{main}) are $\tilde p(a,b) = p(a,b)$, $\Pi^{\tilde A}_a =
P^{\tilde A}_a \otimes I_{\tilde B}$ and $\Pi^{\tilde B}_b = I_{\tilde
A} \otimes P^{\tilde B}_b$.  Different systems \realD\ can generate
the same probabilities $\tilde p(a,b)$.  Let $\hat A$ ($\hat B$) be
the support of the residual density matrix of $\Psi_{\tilde A \tilde
B}$ after a partial trace over $\tilde B$ ($\tilde A$).  The
definition of $P^{\tilde A}_a$ and $P^{\tilde B}_b$ outside $\hat A$
and $\hat B$ respectively cannot affect the probabilities $\tilde
p(a,b)$.  Therefore the only constraint that we can hope to obtain on
$P^{\tilde A}_a$ ($P^{\tilde B}_b$) will have to be a constraint on
$P^{\hat A}_a \stackrel{def}{=} P_{\hat A} P^{\tilde A}_a P_{\hat A}$
($P^{\hat B}_b \stackrel{def}{=} P_{\hat B} P^{\tilde B}_b P_{\hat
B}$), where $P_{\hat A}$ ($P_{\hat B}$) is the projection on the
subspace $\hat A$ ($\hat B$) of $\tilde A$ ($\tilde B$). Our result
takes this fact into account.

Another point is that the defective system $\tilde A$ (the same for
$\tilde B$) might not contain any qubit, that is, there might not
exist any qubit $A$ such that, for some other component $E_A$, we have
$\tilde A = A \otimes E_A$. A unitary transformation on $\tilde A$ can
correspond to a change of basis in $\tilde A$, but it will not create
a qubit because of the trivial fact that the final space is still
$\tilde A$.  If the space $\tilde A$ already had a tensor product
structure, a unitary transformation on this space could change a pure
state into an entangled state, and this could be interpreted as a
modification of the tensor product structure of the state but not of
the space. If the space $\tilde A$ has no tensor product structure, a
unitary transformation on $\tilde A$ will not create one. In
opposition, we would like to conclude that somehow the subspace $\hat
A$ ($\hat B$) of the defective systems $\tilde A$ ($\tilde B$)
contains the correct qubit $A$ ($B$).  One solution is simply to add
an extra qubit $A$ ($B$) initially in the state $|0\rangle^A$
($|0\rangle^B$) and then consider a local unitary transformation
defined on $A \otimes \hat A$ ($B \otimes \hat B$) that will extract
the information about the correct qubits in $\hat A$ ($\hat B$) and
swap it into $A$ ($B$).  Another formal solution is to consider an
isometry from $\hat A$ to $A \otimes E_A$.  An isometry is the same as
an unitary transformation except that it can change the tensor product
structure because the final space is not the same as the initial
space.

\begin{definition}\label{IsoDef}
A linear transformation $U$ from a space $V$ to a space $W$ is an {\em
isometry} if and only if $V$ and $W$ have the same dimension and $U$
preserves the inner product between any pair of states in $V$.
\end{definition}

\begin{theorem} \label{main}
For every setting \realD\ such that $\tilde p(a,b) = p(a,b)$,
$P^{\tilde A}_a$ acts on $\tilde A$ only and $P^{\tilde B}_b$ acts on
$\tilde B$ only, there exists two ``garbage'' spaces $E_A$ and $E_B$,
a state $\Psi_{E_A E_B} \in E_A \otimes E_B$, an isometry $U_{\hat A}$
from $\hat A$ to $A \otimes E_A$ and an isometry $U_{\hat B}$ from
$\hat B$ to $B \otimes E_B$ such that,
\begin{itemize}
\item for every $a \in \Theta$, 
$U_{\hat A} P^{\hat A}_a U_{\hat
A}^\dagger = (P^A_a \otimes I_{E_A})$,
\item for every $b \in \Theta$, $U_{\hat B} P^{\hat B}_b U_{\hat
B}^\dagger = (P^B_b \otimes I_{E_B})$,
\item $(U_{\hat A} \otimes U_{\hat B}) \Psi_{\tilde A \tilde B} =
(\Phi^+_{AB} \otimes \Psi_{E_A E_B})$.
\end{itemize}
\end{theorem} 
This theorem essentially states that the real system is identical to
the ideal system $((P^B_b \otimes I_{E_B}), (P^A_a \otimes I_{E_A}),
\Phi^+_{AB} \otimes \Psi_{E_A E_B})$ up to a local change of basis
(that can modify the tensor product structure) on each side.  The
system $((P^B_b \otimes I_{E_B}), (P^A_a \otimes I_{E_A}), \Phi^+_{AB}
\otimes \Psi_{E_A E_B})$ follows exactly the original specification
except for an additional system $E_A \otimes E_B$ that is in some pure
state $\Psi_{E_A E_B}$ which doesn't interfer at all with this
specification.  One cannot hope to prove more than theorem \ref{main}
using only $\tilde p_a(x) = p_a(x)$. The proof will be given in
section \ref{Proof}.

\section{Connection with the BB84 protocol}  \label{BB84Connection}
There are different ways in which our main result could be connected
to a security proof.  In particular, our self-checking apparatus can
be used in the BB84 protocol in two different ways.  In one way, the
two measuring apparatus in our self-checking apparatus are on Alice's
side.  This is the approach that we will describe here.  It
corresponds to the original idea of a self-checking source entirely
located on Alice's side.  The other approach is that the two measuring
apparatus are respectively located on Alice's side and on Bob's side.
We will not discuss this other option here.

We recall that we consider a variation on the BB84 protocol where
every state is rotated of an angle $-\pi/8$ so that the bases used are
at angle $-\pi/8$ and $\pi/8$ instead of $0$ and $\pi/4$ as is usually
the case in the literature.  When Alice picks $\alpha
=\{-\pi/8,\pi/8\}$ for the measurement on the first photon and obtains
the outcome $x$, the second photon collapses into the state $|\alpha +
x \pi/2\rangle$, as requested in the BB84 protocol.  However, our test
requires that Alice uses the three angles $-\pi/8$, $0$ and $\pi/8$
for the two photons.  Therefore, to test the apparatus Alice will pick
a random set of positions $R$ and the bases $\alpha, \beta \in
\{-\pi/8, 0, \pi/8\}$ for the positions in this set.  For the non
tested positions (i.e. the positions not in $R$), Alice will use
$\alpha \in \{-\pi/8, \pi/8\}$ for the first photon and send the other
photon to Bob.  The basic intuition is that, if the test really works,
it should be unlikely that this test succeeds on $R$ and would fail if
it was executed outside $R$.

Let us prove that, whenever the source respect the conclusion of
theorem \ref{main}, the protocol is as secure as if an ordinary BB84
source was used.  It is convenient to consider $U_{\hat A} \otimes
U_{\hat B}$ as a change of bases which provides an alternative
representation for the states of the subsystem $\hat A \otimes \hat
B$.  In this alternative representation, it is not hard to see that if
$(\alpha, x)$ is used/obtained on $A$'s side, the system $B \otimes
E_A \otimes E_B$ must be left in the collapsed state $| \alpha + x
\pi/2\rangle \otimes \Psi_{E_A E_B}$.  If we return to the original
representation, the collapsed state is $(U_{\hat A} \otimes U_{\hat
B}) |\alpha + x \pi/2\rangle^A \otimes | \alpha + x \pi/2\rangle^B
\otimes \Psi_{E_A E_B}$.  The transformation $U_{\hat A}$ has no
effect on $\tilde B$'s side, so it's the same thing as if Eve received
the part $B \otimes E_B$ of $U_{\hat B} |{a_1} + x_1
\pi/2\rangle^{H_2} \otimes \Psi_{E_A E_B}$.  An important fact is that
$U_{\hat B}$ and $\Psi_{E_A E_B}$ are independent of $\alpha$ and $x$.
Therefore, with the state $|\alpha + x \pi/2\rangle^B$ and the part
$E_B$ of the state $\Psi_{E_A E_B}$ Eve could herself create the state
$U_{\hat B} | \alpha + x \pi/2\rangle^B \otimes \Psi_{E_A
E_B}$. Therefore, Eve has nothing more than what she could obtain if
the ordinary BB84 source was used together with a completely
uncorrelated state $\Psi_{E_A E_B}$ that is initially shared between
Alice and Bob.

\section{The proof}\label{Proof}
Here we prove the main result (theorem \ref{main}). Theorem \ref{main}
is given in terms of two local isometries $U_{\hat A}$ and $U_{\hat
B}$ which preserve the tensor product structure of the subspace $\hat
A \otimes \hat B$ of $\tilde A \otimes \tilde B$.  We will also need a
simpler notion of isomorphism which ignores the tensor product
structure of the space $\tilde A \otimes \tilde B$.  If we do not care
about the tensor product structure, there is a smaller space which
contains $\Psi_{\tilde A \tilde B}$ and this space is sufficient to
describe the essential of the projections $P^{\tilde A}_a$ and
$P^{\tilde B}_b$. This space is the span $S$ of $\{ (P^{\tilde A}_a
\otimes P^{\tilde B}_b ) \Psi_{\tilde A \tilde B} \;|\; a,b \in \Theta
\}$.
\begin{definition} \label{InnerProductIso}
Consider any setting \realP.  The setting \realP\ is inner product
isomorph to \spec\ if there exists an isometry $U$ from the span $S$
of $\{ (P^{\tilde A}_a \otimes P^{\tilde B}_b ) \Psi_{\tilde A \tilde
B} \;|\; a,b \in \Theta \}$ to $A \otimes B$ such that, for every $a,
b \in \Theta$, for every $|\phi_S\rangle \in S$, we have
\begin{itemize}
\item[A1:] $(P^{\tilde A}_a \otimes P^{\tilde B}_b) |\phi_S\rangle =
  U^{\dagger} (P^A_a \otimes P^B_b) U |\phi_S\rangle$ and
\item[A2:] $U \Psi_{\tilde A \tilde B} = \Phi^+_{AB}$.
\end{itemize}
\end{definition}
The proof of theorem \ref{main} proceeds in two main steps.  First, we
prove that the equality $\tilde p(a,b) = p(a,b)$ implies that \realP\
is inner product isomorph to \spec.  Second, we show that this
isomorphism implies the conclusion of theorem \ref{main}. The reader
might find the second step a little bit surprising because $S$, the
span of $\{ (P^{\tilde A}_a \otimes P^{\tilde B}_b) \Psi_{\tilde A
\tilde B} \;|\; a,b \in \Theta \}$, is not necessarily identical to
$\hat A \otimes \hat B$. In fact, $S$ is not in general the tensor
product of two Hilbert spaces.  As we will see, the trick is that the
inner product structure $\hat A \otimes \hat B$ can be reconstructed
because the projections $P^{\tilde A}_a$ and $P^{\tilde B}_b$ are
defined on $\tilde A$ and $\tilde B$ separately.

\subsection{The inner product isomorphism.}  
Throughout this subsection we assume that the equality $\tilde p(a,b)
= p(a,b)$ hold and we try to show that \realP\ is inner product
isomorph to \spec.  We have $6$ possible values for $a$ (3 bases with
2 outcomes each) and $6$ possible values for $b$, so a total of 36
possible (non normalised) states $(P^{\tilde A}_a \otimes P^{\tilde
B}_b) \Psi_{\tilde A \tilde B}$.  If our goal can be achieved, these
36 vectors should lie in a 4 dimensional space and be linearly related
as in the ideal specification.  For every $\alpha \in \{-\pi/8, 0,
\pi/8\}$, let $\Theta_\alpha = \{ (\alpha, 0), (\alpha, 1)\}$. To
achieve our goal we have that the lenght of these vectors are uniquely
determined by the probabilities $p(a,b)$. We also have that, for every
$\alpha \in \{-\pi/8, 0, \pi/8\}$, $\sum_{a\in \Theta_{\alpha}}
P^{\tilde A}_a = I$, $\sum_{b \in \Theta_{\alpha}} P^{\tilde B}_b =
I$, and also the commutativity of $P^{\tilde A}_a$ and $P^{\tilde
B}_b$.  The proof of the inner product isomorphism proceeds in three
steps. In the first step, we show the following simple proposition.
\begin{proposition} \label{prop1}
For every $a \in \Theta$,
\[
P^{\tilde A}_a P^{\tilde B}_a
\Psi_{\tilde A \tilde B} = P^{\tilde B}_a P^{\tilde A}_a \Psi_{\tilde
A \tilde B} = 
P^{\tilde A}_a \Psi_{\tilde A \tilde B} =
 P^{\tilde B}_a
\Psi_{\tilde A \tilde B}.
\]
\end{proposition}
In the second step, we show this other simple proposition.
\begin{proposition} \label{prop2}
For every $(\alpha, \beta) \in \{-\pi/8, 0, \pi/8\}$ with $\alpha \neq
\beta$, the four non normalised vectors in ${\cal B}_{(\alpha, \beta)}
\stackrel{def}{=} \{ P^{\tilde A}_a P^{\tilde B}_b \Psi_{\tilde A
\tilde B}\; |\; a \in \Theta_{\alpha}, b \in \Theta_{\beta} \}$, are
orthogonal and have the same length as the corresponding ideal vectors
$P^{A}_a P^{B}_b \Psi^+_{AB}$.
\end{proposition}
Each of these 6 different sets ${\cal B}_{(\alpha, \beta)}$ of 4
vectors span a 4 dimensional space.  The third and crucial step is to
show that, for every $(\alpha, \beta)$ and $(\alpha', \beta')$ with
$\alpha \neq \beta$ and $\alpha' \neq \beta'$, 
the four vectors in
${\cal B}_{(\alpha, \beta)}$ are linearly related to the four vectors
in ${\cal B}_{\alpha', \beta'}$ with the same coefficients as for the
corresponding vectors in the ideal specification. More precisely, we
must show the following crucial proposition.
\begin{proposition} \label{prop3}
For every $(\alpha, \beta)$ and $(\alpha', \beta')$ with $\alpha \neq
\beta$ and $\alpha' \neq \beta'$, for every $(x,y) \in \{0,1\}^2$,
\begin{eqnarray*}
\lefteqn{P^{\tilde A}_{(\alpha,x)} P^{\tilde B}_{(\beta,y)}
\psi_{\tilde A \tilde B}} \nonumber \\ & = & \sum_{(x',y') \in
\{0,1\}} [T_{(\alpha,\beta)}^{(\alpha', \beta')}]^{(x',y')}_{(x,y)}
P^{\tilde A}_{(\alpha',x')} P^{\tilde B}_{(\beta',y')} \psi_{\tilde A
\tilde B}
\end{eqnarray*}
where $[T_{(\alpha,\beta)}^{(\alpha', \beta')}]$ is the unique $4
\times 4$ coefficient matrix such that 
\begin{eqnarray*}
\lefteqn{P^{A}_{(\alpha,x)} P^{B}_{(\beta,y)} \Phi^+} \\ & = &
\sum_{(x',y') \in \{0,1\}} [T_{(\alpha,\beta)}^{(\alpha',
\beta')}]^{(x',y')}_{(x,y)} P^{A}_{(\alpha',x')} P^{B}_{(\beta',y')}
\Phi^+_{AB}.
\end{eqnarray*}
\end{proposition}
It is not hard to see that proposition \ref{prop3} implies that the
mapping that maps $P^{\tilde A}_a P^{\tilde B}_b \Psi_{\tilde A \tilde
B}$ into $P^A_a P^B_b \Phi^+_{AB}$, for $a, b \in \Theta_0 =
\{(0,0),(0,1)\}$, is an inner product isomorphism (see definition
\ref{InnerProductIso}).  Indeed, proposition \ref{prop3} implies that
any of the 6 sets ${\cal B}_{(\alpha,\beta)}$ is a (non normalised)
basis for $S$ that is as good as any other basis to check if a
projection $P^{\tilde A}_a P^{\tilde B}_b$ has the correct matrix
representation. So, to check $P^{\tilde A}_a P^{\tilde B}_b$, one
simply picks the basis ${\cal B}_{(\alpha,\beta)}$ that contains
$P^{\tilde A}_a P^{\tilde B}_b \Psi_{\tilde A \tilde B}$.

Note that proposition \ref{prop2} consider 24 vectors, and we said
that there are 36 outcomes.  This can be explained by two facts.
First, the probabilities are the same in the real setting as in the
ideal setting and, thus, 6 of these vectors vanish because their
associated outcome occurs with probability zero: for every $\alpha \in
\{-\pi/8, 0, \pi/8\}$
\begin{equation}\label{eq1}
P^{\tilde A}_{(\alpha,0)} P^{\tilde B}_{(\alpha,1)} \Psi_{\tilde A
\tilde B} = P^{\tilde A}_{(\alpha,1)} P^{\tilde B}_{(\alpha,0)}
\Psi_{\tilde A \tilde B} = 0.
\end{equation}
This brings us back to 30 vectors.  Second, proposition \ref{prop1}
says that the 6 non vanishing vectors $P^{\tilde A}_a P^{\tilde B}_a
\Psi_{\tilde A \tilde B} = P^{\tilde B}_a P^{\tilde A}_a \Psi_{\tilde
A \tilde B}$, $a \in \Theta$, can be written as $P^{\tilde A}_a
\Psi_{\tilde A \tilde B}$ or $P^{\tilde B}_a \Psi_{\tilde A \tilde B}$
and thus are known linear combinations of the 24 vectors considered in
proposition \ref{prop2}.  Now, we prove proposition~\ref{prop1}.
\debproof{ of proposition \ref{prop1}} We have $P^{\tilde A}_a
P^{\tilde B}_a \Psi_{\tilde A \tilde B} = P^{\tilde B}_a \Psi_{\tilde
A \tilde B}$ because the collapse associated with the projection
$P^{\tilde A}_a$ on $P^{\tilde B}_a \Psi_{\tilde A \tilde B}$ occurs
with probability $1$.  Similarly, $P^{\tilde B}_a P^{\tilde A}_a
\Psi_{\tilde A \tilde B} = P^{\tilde A}_a \Psi_{\tilde A \tilde B}$.
By commutativity, we obtain $P^{\tilde B}_a P^{\tilde A}_a
\Psi_{\tilde A \tilde B} = P^{\tilde A}_a P^{\tilde B}_a \Psi_{\tilde
A \tilde B}$.  By transitivity, we obtain proposition \ref{prop1}b.
This concludes the proof.  \finproof
\debproof{ of proposition \ref{prop2}} The orthogonality of the four
vectors $P^{\tilde A}_a P^{\tilde B}_b \Psi_{\tilde A \tilde B}$, $a
\in \Theta_{\alpha}$ and $b \in \Theta_{\beta}$, is immediate from the
fact that $P^{\tilde A}_{(\alpha,0)}$ is orthogonal to $P^{\tilde
A}_{(\alpha,1)}$, $P^{\tilde B}_{(\beta,0)}$ is orthogonal to
$P^{\tilde B}_{(\beta,1)}$ and the commutativity of $P^{\tilde A}_a$
and $P^{\tilde B}_b$.  The length of the vector $P^{\tilde A}_a
P^{\tilde B}_b \Psi_{\tilde A \tilde B}$ is the same as in the ideal
specification because it is uniquely determined by the probability
$\tilde p(a,b) = p(a,b)$. \finproof 
Now, we proceed with the third step which we feel is the most
importany step of the entire proof.
\debproof{ of proposition \ref{prop3}} Here is a crucial observation
for the proof. Let $(\alpha, \gamma, \beta)$ be either $(-\pi/8, 0,
\pi/8)$, $(0, \pi/8, -\pi/8)$ or $(\pi/8, -\pi/8, 0)$: a cyclic
permutation of $(-\pi/8, 0, \pi/8)$.  In this way, the value of $\beta
\in \{-\pi/8, 0, \pi/8\}$ uniquely determines $\alpha$ and $\gamma$.
In the specified setting, if we use a fixed basis $\beta$ and a fixed
outcome $z$ on one side, say the side $\tilde B$, and look at the
different final states associated with the two different bases
$\alpha$ and $\gamma$ and the two different outcomes $0$ and $1$ on
the other side, we see that the 4 different final states lie in the
same real two dimensional plane.  This fact is easily understood
because the system on the non fixed side is a two dimensional system
and the states in the measurement bases are all in the same real plane
(no complex numbers). Two of these four states belong to ${\cal
B}_{(\alpha,\beta)}$ whereas the two others belong to ${\cal
B}_{(\gamma,\beta)}$.
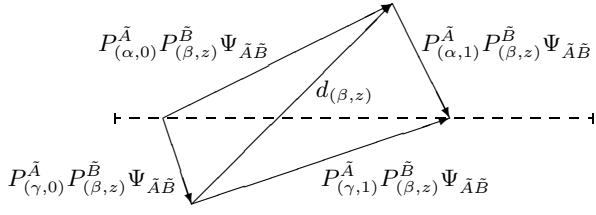
\begin{figure}[p]
\setlength{\unitlength}{.05in}
\begin{center}
\begin{picture}(50,25)(0,0)

\blue{ \put(10,10){\vector(2,1){24}}
\put(22,16){\makebox(0,0)[rb]{$P^{\tilde A}_{(\alpha,0)} P^{\tilde
B}_{(\beta,z)} \Psi_{\tilde A \tilde B}$ }}
\put(34,22){\vector(1,-2){6}} 
\put(37,16){\makebox(0,0)[lb]{$P^{\tilde
A}_{(\alpha,1)} P^{\tilde B}_{(\beta,z)} \Psi_{\tilde A \tilde B}$}}}

\put(5,10){\dashbox{1}(50,0){}}

\green{ 
\put(10,10){\vector(1,-3){3}} 
\put(11.5,5.5)%
{\makebox(0,0)[rt]{$P^{\tilde A}_{(\gamma,0)} P^{\tilde B}_{(\beta,z)}
\Psi_{\tilde A \tilde B}$}} 
\put(13,1){\vector(3,1){27}}}
\put(26.5,5.5)%
{\makebox(0,0)[lt]{$P^{\tilde A}_{(\gamma, 1)} P^{\tilde B}_{(\beta,z)} 
\Psi_{\tilde A \tilde B}$ }}

\put(13,1){\vector(1,1){21}}
\put(26,14){\makebox(0,0)[lt]{$d_{(\beta,z)}$}}
\end{picture}
\end{center}
\caption{The vector $d_{(\beta,z)}$ has maximum length when the real
plane spanned by $P^{\tilde A}_{(\alpha,0)} P^{\tilde B}_{(\beta,z)}
\Psi_{\tilde A \tilde B}$ and $P^{\tilde A}_{(\alpha,1)} P^{\tilde
B}_{(\beta,z)} \Psi_{\tilde A \tilde B}$ (above the dotted line) and
the real plane spanned by $P^{\tilde A}_{(\gamma,0)} P^{\tilde
B}_{(\beta,z)} \Psi_{\tilde A \tilde B}$ and $P^{\tilde A}_{(\gamma,
1)} P^{\tilde B}_{(\beta,z)} \Psi_{\tilde A \tilde B}$ (below the
dotted line) are one and the same plane.}
\label{figure}
\end{figure}

Note that states that have different value of $z$ are
orthogonal. So, the linear relationship between ${\cal
B}_{(\alpha,\beta)}$ and ${\cal B}_{(\gamma,\beta)}$ does not mix
different values of $z$: the coefficient
$[T_{(\gamma,\beta)}^{(\alpha, \beta)}]^{(x',z')}_{(x,z)}$ vanishes
when $z \neq z'$. Therefore, it is sufficient to see, for each value
of $z$ individually, how the two states associated with $(\gamma,
\beta)$ and $z$ (below the dotted line in figure \ref{figure}) are
linearly related to the two states associated $(\alpha, \beta)$ and
$z$ (above the dotted line in figure \ref{figure}). For $z = 0,1$, let
\[
d_{(\beta,z)} = ( P^{\tilde A}_{(\alpha,0)} P^{\tilde B}_{(\beta, z)}
- P^{\tilde A}_{(\gamma,0)} P^{\tilde B}_{(\beta, z)}) \Psi_{\tilde A
\tilde B}.
\]  
It is not hard to algebrically check, with the help of proposition
\ref{prop1}, that the length of
\[
d_\beta \stackrel{def}{=} d_{(\beta,0)} + d_{(\beta,1)} = 
(P^{\tilde
A}_{(\alpha,0)} - P^{\tilde A}_{(\gamma,0)} )\Psi_{\tilde A \tilde B}
\] 
is uniquely determined by the probabilities $\tilde p(a,b) =
p(a,b)$. Indeed, we have
\begin{eqnarray*}
&& \langle \Psi_{\tilde A \tilde B}| (P^{\tilde A}_{(\alpha,0)} -
P^{\tilde A}_{(\gamma,0)} ) (P^{\tilde A}_{(\alpha,0)} - P^{\tilde
A}_{(\gamma,0)} )| \Psi_{\tilde A \tilde B}\rangle\\ && \quad =
\|P^{\tilde A}_{(\alpha,0)} | \Psi_{\tilde A \tilde B}\rangle\|^2 -
\langle \Psi_{\tilde A \tilde B}| P^{\tilde A}_{(\gamma,0)} P^{\tilde
A}_{(\alpha,0)} | \Psi_{\tilde A \tilde B}\rangle \\ && \quad \quad -
\langle \Psi_{\tilde A \tilde B}| P^{\tilde A}_{(\alpha,0)} P^{\tilde
A}_{(\gamma,0)} | \Psi_{\tilde A \tilde B}\rangle + \| P^{\tilde
A}_{(\gamma,0)}| \Psi_{\tilde A \tilde B} \rangle \|^2
\end{eqnarray*}
and
\begin{eqnarray*}
\langle \Psi_{\tilde A \tilde B}| P^{\tilde A}_{(\gamma,0)} P^{\tilde
A}_{(\alpha,0)} | \Psi_{\tilde A \tilde B}\rangle & = & \langle
\Psi_{\tilde A \tilde B}| P^{\tilde A}_{(\gamma,0)} P^{\tilde
B}_{(\alpha,0)} | \Psi_{\tilde A \tilde B}\rangle \\ & = & \|
P^{\tilde A}_{(\gamma,0)} P^{\tilde B}_{(\alpha,0)} | \Psi_{\tilde A
\tilde B}\rangle \|^2,
\end{eqnarray*}
and similarly for the term $\langle \Psi_{\tilde A \tilde B}|
P^{\tilde A}_{(\alpha,0)} P^{\tilde A}_{(\gamma,0)} | \Psi_{\tilde A
\tilde B}\rangle$.  We also have that the two states $d_{(\beta,0)}$
and $d_{(\beta,1)}$ are orthogonal because they have different value
of $z$.  So, we have obtained
\begin{eqnarray}
\| d_{(\beta,0)} \|^2 + \| d_{(\beta,1)} \|^2 & = & \| d_\beta \|^2
\nonumber \\ 
& = & \| d_\beta^{ideal} \|^2 \nonumber \\ 
& = & \| d_{(\beta,0)}^{ideal} \|^2 + \| d_{(\beta,1)}^{ideal} \|^2.
 \label{FirstConstraint}
\end{eqnarray}
After an exhaustive consideration of all cases, one can check that in
the ideal specification we have that either, for both $z = 0$ and $z=
1$, $P^{A}_{(\alpha,0)} P^{B}_{(\beta,z)} \Phi^+$ is on the same side
of the dotted line (see figure \ref{figure}) as $P^{A}_{(\gamma,0)}
P^{B}_{(\beta,z)} \Phi^+_{AB}$ or else, for both $z = 0$ and $z = 1$,
$P^{A}_{(\alpha,0)} P^{B}_{(\beta,z)} \Phi^+$ is on a different side
of the dotted line than $P^{A}_{(\gamma,0)} P^{B}_{(\beta,z)}
\Phi^+_{AB}$. This means that either both $\| d_{(\beta,0)}^{ideal}
\|^2$ and $\| d_{(\beta,1)}^{ideal} \|^2$ reach their maximum value or
else they both reach their minimum value. So, we have the following.
\begin{eqnarray} 
& \| d_{(\beta,z)} \|^2 \leq \| d_{(\beta,0)}^{ideal} \|^2 \quad (\forall z
\in \{0,1\})&  \nonumber  \\
& \mbox{or} &  \label{SecondConstraint} \\
& \| d_{(\beta,z)} \|^2 \geq \| d_{(\beta,0)}^{ideal} \|^2 \quad (\forall z
\in \{0,1\}) & \nonumber 
\end{eqnarray}
If we combine (\ref{FirstConstraint}) and (\ref{SecondConstraint}), we
obtain $\| d_{(\beta,0)} \|^2 = \|d_{(\beta,0)}^{ideal} \|^2$ and $\|
d_{(\beta,1)} \|^2 = \|d_{(\beta,1)}^{ideal} \|^2$. So, the real
setting reaches the same extreme situation as in the ideal setting
where the coefficients $[T_{(\gamma,\beta)}^{(\alpha,
\beta)}]^{(x',z')}_{(x,z)}$ are uniquely determined. This shows that
the transformation $[T_{(\gamma,\beta)}^{(\alpha,
\beta)}]^{(x',z')}_{(x,z)}$ is the same as in the ideal case.

Now, we must consider arbitrary transformation from ${\cal
B}_{(\alpha, \beta)}$ to ${\cal B}_{(\alpha', \beta')}$, not just from
${\cal B}_{(\alpha, \beta)}$ to ${\cal B}_{(\gamma, \beta)}$.  By
symmetry, we also have the transformations from ${\cal
B}_{(\alpha,\beta)}$ to ${\cal B}_{(\alpha,\gamma)}$, that is, we can
also change the basis on the side $\tilde B$ while keeping the same
basis on the side $\tilde A$. For any $(\alpha,\beta)$,
$(\alpha',\beta')$ and $(\alpha'',\beta'')$, the transformation from
${\cal B}_{(\alpha, \beta)}$ to ${\cal B}_{(\alpha'', \beta'')}$ is
the product of the transformation from ${\cal B}_{(\alpha, \beta)}$ to
${\cal B}_{(\alpha', \beta')}$ with the transformation from ${\cal
B}_{(\alpha', \beta')}$ to ${\cal B}_{(\alpha'', \beta'')}$.  Using
this fact, it is easy to obtain all transformations from an arbitrary
set ${\cal B}_{(\alpha,\beta)}$ to an arbitrary set ${\cal
B}_{(\alpha', \beta')}$.  This concludes the proof. \finproof

\subsection{The tensor product structure}
Here, we use the result of the previous subsection to prove theorem
\ref{main}.  We want to construct two local isometries $U_{\hat A}$
and $U_{\hat B}$ that respects the three conditions of theorem
\ref{main}.  Intuitively, the isometry $U_{\hat A}$ will extract the
information about the correct qubit that is hidden inside $\hat A$,
and leave $\hat A$ (without this information) as the garbage space. It
will swap the state of a correct qubit that is somehow hidden in $\hat
A$ with the state of an additional qubit $A$ that is initially in
state $|0\rangle$. This is just an intuition.

Formally, we will only show how to obtain $U_{\hat A}$ because
$U_{\hat B}$ can be obtained in the same way with an additional qubit
$B$ that is initially in the state $|0\rangle^{B}$.  For two qubits
$A'$ and $A$, let us denote $\stackrel{\rightarrow}{N}_{AA'}$ the
control not operation where $A$ is the source qubit and $A'$ the
target qubit. Similarly, let us denote
$\stackrel{\leftarrow}{N}_{AA'}$ the control not operation where $A'$
is the source and $A$ is the target. Let $N_A$ and $N_{A'}$ be the
not operation on $A$ and $A'$, respectively. One can easily check that
\begin{equation} \label{forwardCNOT}
\stackrel{\rightarrow}{N}_{A A'} = P^{A}_{(0,0)} \otimes I_{A'} +
P^{A}_{(0,1)} \otimes N_{A'}
\end{equation}
\begin{equation}  \label{backwardCNOT}
\stackrel{\leftarrow}{N}_{A A'} = I_{A} \otimes P^{A'}_{(0,0)} + N_{A}
\otimes P^{A'}_{(0,1)},
\end{equation}
and
\begin{equation}\label{NOT}
N_{A/A'} = \sqrt 2(P^{A/A'}_{(\pi/8,0)} - P^{A/A'}_{(-\pi/8,0)}).
\end{equation}
Note that $\stackrel{\rightarrow}{N}_{ A A'}
\stackrel{\leftarrow}{N}_{A A'}$ is the standard swap operation on $A
\otimes A'$ given that $A$ is initially in the state $|0\rangle^A$.
This suggests that we define the swap operation on $A \otimes \hat A$
as
\begin{equation}\label{UhatDefinition}
U_{A \hat A} \stackrel{def}{=} \stackrel{\rightarrow}{N}_{A \hat A}
\stackrel{\leftarrow}{N}_{A \hat A}
\end{equation}
where the extended control not operations
$\stackrel{\rightarrow}{N}_{A \hat A}$ and
$\stackrel{\leftarrow}{N}_{A \hat A}$, by analogy with
(\ref{forwardCNOT}), (\ref{backwardCNOT}) and (\ref{NOT}), are defined
as
\begin{equation}\label{defforwardCNOT}
\stackrel{\rightarrow}{N}_{A \hat A} \stackrel{def}{=} P^{A}_{(0,0)}
\otimes I_{\hat A} + P^{A}_{(0,1)} \otimes N_{\hat A}
\end{equation}
and
\begin{equation}\label{defbackwardCNOT}
\stackrel{\leftarrow}{N}_{A \hat A} \stackrel{def}{=} I_{A} \otimes
P^{\hat A}_{(0,0)} + N_{A} \otimes P^{\tilde A}_{(0,1)}
\end{equation}
where
\[
N_{\hat A} \stackrel{def}{=} \sqrt 2(P^{\hat A}_{(\pi/8,0)} -
P^{\hat A}_{(-\pi/8,0)}).
\]
The swap operation $U_{B \hat B}$ can be defined in a similar way.

We recall that the span of $\{ (P^{\tilde A}_a \otimes P^{\tilde B}_b)
\Psi_{\tilde A \tilde B} \;|\; a,b \in \Theta \}$ is denoted as $S$.
Let us show that that, for every $\Phi_S \in S$, for every
$|a\rangle^A \in A$, we have
\begin{eqnarray} 
&& (U_{A \hat A} \otimes I_{\hat B}) (|a\rangle^A \otimes \Phi_S)
\label{swap} \\
&& \quad =(I_{A} \otimes U^\dagger) 
(\stackrel{\rightarrow}{N}_{A A'}
\stackrel{\leftarrow}{N}_{A A'} \otimes I_{B'}) (I_{A} \otimes U)
(|a\rangle^A \otimes \Phi_S) \nonumber
\end{eqnarray}
where $U$ is the inner product isomorphism from $S$ to $A' \otimes
B'$.  First, we use the definitions of $U_{A \hat A}$,
$\stackrel{\rightarrow}{N}_{\tilde A A}$ and
$\stackrel{\leftarrow}{N}_{\tilde A A}$ respectively given in
(\ref{UhatDefinition}), (\ref{defforwardCNOT}) and
(\ref{defbackwardCNOT}) to expand $U_{A \hat A}$ in the LHS in terms
of the projections $P^{\hat A}_a$ and $P^A_a$.  Second, use the
condition $A1$ of definition \ref{InnerProductIso} for an inner
product isomorphism (summing over $b \in \Theta_0$ to get rid of the
projections $P^{\hat B}_b$) to replace every projection $P^{\hat A}_a$
by its corresponding projection $P^{A'}_a$. This also adds $(I_{A}
\otimes U^\dagger)$ to the left and $(I_{A} \otimes U)$ to the right.
To finally obtain the RHS, we use the identities (\ref{forwardCNOT}),
(\ref{backwardCNOT}) and (\ref{NOT}) to recover the expression
$(\stackrel{\rightarrow}{N}_{AA'} \stackrel{\leftarrow}{N}_{AA'}
\otimes I_{B'})$, but this time on $A \otimes A' \otimes B'$. We can
obtain a similar result for $U_{B \hat B}$.

With the help of (\ref{swap}) and condition $A2$ of the inner product
isomorphism, we can obtain
\begin{equation}\label{ThirdCondition}
(U_{A \hat A} \otimes U_{B \hat B}) |0\rangle^A \otimes |0\rangle^B
\otimes \Psi_{\tilde A \tilde B} = \Phi^+_{AB} \otimes \Psi_{E_A E_B}
\end{equation}
with
\[
\Psi_{E_A E_B} = U^\dagger (|0\rangle^{A'} \otimes |0\rangle^{B'}).
\] 
This is essentially the third condition of theorem \ref{main}.  It is
not exactly the third condition because it is expressed in terms of
the swap operation $U_{A \hat A}$ and $U_{B \hat B}$, not in terms of
isometries.  The isometry $U_{\hat A}$ is simply the transformation
$U_{A \hat A}$ where the space $A$ is always in the fixed state
$|0\rangle^A$ so that the fixed component $|0\rangle^A$ does not need
to appear explicitly in the initial state.  The image of $U_{\hat A}$
on $\hat A$ is the image of $U_{A \hat A}$ on $\{|0\rangle^A\} \otimes
\hat A$.  It is not hard to see using (\ref{ThirdCondition}) and
proposition \ref{prop4} (provided later) that the image of $U_{\hat
A}$ is $A \otimes E_A$ where $E_A$ is the support of the residual
density matrix of $\Psi_{E_A E_B}$ on $\hat A$.  The isometry $U_{\hat
B}$ can be defined in a similar way.  With these definitions,
(\ref{ThirdCondition}) becomes
\[
(U_{\hat A} \otimes U_{\hat B}) \Psi_{\tilde A \tilde B} = \Phi^+_{AB}
\otimes \Psi_{E_A E_B}
\]
which is exactly the third condition. With the help of (\ref{swap}),
as a first step toward the first condition, we can obtain that, for
every $\Phi_S \in S \subseteq \hat A \otimes \hat B$,
\begin{eqnarray} 
&& (P^{\hat A}_a \otimes I_{\hat B}) \Phi_S  \nonumber \\ 
&& \quad  = U_{\hat A}^\dagger (P^A_a \otimes I_{\hat A
\hat B}) U_{\hat A} \Phi_S 
\label{FirstCondOnS}
\end{eqnarray}
and similarly for the second condition for $P^{\hat B}_b$.

To conclude the proof, it only remains to show that $(U_{\hat A}
\otimes I_{\hat B})$ does the right job on $\hat A \otimes \hat B$, not
only on $S$ (i.e., $U_{\hat A}$ is an isometry and fully satisfied the
first condition of the theorem), and similarly with $U_{\hat B}$ (for
the second condition).  The following easy to prove proposition is a
key ingredient.

\begin{proposition} \label{prop4}
Let $\Psi_{XY}$ be any entangled pure state of a bipartite system $X
\otimes Y$.  Let $\hat X$ be the support of the residual density
matrix of $\Psi_{XY}$ on $X$. Let $A_{X}$ and $A'_{X}$ be any two
linear operators from $\hat X$ to another space $Z$.  We have that
$A_X \Psi_{XY} = A'_X \Psi_{XY}$ if and only if $A_X =
A'_X$. Moreover, the image of $A_X$ on $\hat X$ is the support of the
residual density matrix of $A_X \Psi_{XY}$ on $Z$.
\end{proposition}

\debproof{} Consider a Schmidt decomposition
\[
\Psi_{XY} = \sum_{i \in {\cal X}} \lambda_i |i\rangle^X \otimes
|i\rangle^Y
\]
of $\Psi_{XY}$ where $\lambda_i > 0$ for every $i \in {\cal X}$.
Consider the projection $P^{Y}_i = |i\rangle^Y\!\!\langle i|$ on
$Y$. We have that $A_X \Psi_{XY} = A'_X \Psi_{XY}$, if and only if,
for every $i \in {\cal X}$, $P^Y_i A_X \Psi_{XY} = P^Y_i A'_X
\Psi_{XY}$ which implies that, for every $i \in {\cal X}$, $A_X
|i\rangle^X \otimes |i\rangle^Y = A'_X |i\rangle^X \otimes
|i\rangle^Y$, and thus $A_X |i\rangle^X = A'_X |i\rangle^X$.  The
image of $A_X$ on $\hat X$ is the span of $\{ A_X |i\rangle^X\; | \; i
\in {\cal X}\}$, but this is also the support of the residual density
matrix of $A_X \Psi_{XY}$ on $Z$.  \finproof

We must show that $U_{\hat A}$ is an isometry from $\hat A$ to its
image (which we do not need to know here), and then check that this
isometry respects the first condition of theorem \ref{main}.  To
obtain that $U_{\hat A}$ is an isometry, it is sufficient to obtain
that $U_{\hat A}^\dagger U_{\hat A}$ is the identity on $\hat A$. We
can use the definition of $U_{\hat A}$ in terms of the projections
$P^{\hat A}_a$ and the inner product isomorphism on $S$ (as we did to
obtain (\ref{swap})), but this time to obtain that $U_{\hat A}^\dagger
U_{\hat A} \Psi_{\tilde A \tilde B} = \Psi_{\tilde A \tilde B}$ and
then apply proposition \ref{prop4} to conclude that $U_{\hat
A}^\dagger U_{\hat A}$ is the identity on $\hat A$.  Now, we verify
the first condition of theorem \ref{main} in a similar way.  We want
to show that
\[
P^{\hat A}_a = U_{\hat A}^\dagger (P^A_a \otimes I_{E_A})
U_{\hat A}
\] 
on $\hat A$.  We simply use (\ref{FirstCondOnS}) with $\Phi_S =
\Psi_{\tilde A \tilde B}$ and proposition \ref{prop4}. 
This concludes the proof of theorem \ref{main}.

\section{Discussion}
Thus far, this result was not generalised to a large class of
settings. Of course, if we only vary the angles a little, it will
still hold, but nothing is known for more interesting variations.  For
example, it is still an open question whether a similar result applies
if we used the angles $\{0, \pi/4\}$ on one side and the angles
$\{\pi/8, -\pi/8\}$ on the other side, instead of $\{\pi/8, 0,
-\pi/8\}$ on both sides.  Also, it is known that the GHZ state can
also be self tested \cite{mayers00}, but no rule is known for a large
class of states.

The theorem was obtained in the context of quantum cryptography. It
would be useful to obtain a robust variation on this theorem in which
the probabilities do not have to exactly respect the ideal
specification.  The idea is that a statistical test could then be used
to obtain enough constraints on the source for the purpose of quantum
cryptography.  One may also ask what are the implications of this
result in the foundation of quantum mechanics.  Somehow, the essential
processus of science is that we try to figure out what specific models
describe the observed classical data in different settings.  In
particular, we make hypothesis and then conduct experiments to verify
them.  This is essentially the situation that is analysed in our
theorem.  Therefore, it would be interesting to see if the analysis of
self-checking quantum apparatus says anything interesting about the
way we build our different quantum models.  If there is a link, it is
not a trivial link because, typically, experimentalits know what their
measuring apparatus are when they verify a model whereas our theorem
applies when the measuring apparatus on each side are not known.
Finally, in view of the fact that configurations such as the one we
consider in our theorem have been studied for many years in the
context of violation of Bell's inequalities, it is interesting to know
that they are uniquely determined (up to a natural isomorphism) by the
probability distributions that they generate.

The first author would like to acknowledge useful discussions with
Daniel Gottesman. This research was done in part while the first
author was professor at Maharishi University of Management and, later,
a visitor at the Perimeter Institute.  It was supported in part by the
National Science Foundation through Caltech's Institute for Quantum
Information under Grant No.  EIA-0086038.

\end{document}